\input harvmac

\def\curl{\nabla\times}

\def\perpp{{\scriptscriptstyle\perp}}

\def\half{{1\over 2}}

\def\dot{\!\cdot\!}

\def\vb{{\bf v}}
\def\vA{{\bf A}}
\def\vH{{\bf H}}
\def\vm{{\bf m}}

\def\bold#1{\setbox0=\hbox{$#1$}%
     \kern-.010em\copy0\kern-\wd0
     \kern.025em\copy0\kern-\wd0
     \kern-.020em\raise.0200em\box0 }
\def\cross{\!\times\!}
\def\grad{\nabla_{\scriptscriptstyle \perp}}

\lref\KNMoire{R.D.~Kamien and D.R.~Nelson, Phys. Rev. Lett. {\bf 74} (1995)
2499 [cond-mat/9411039]; Phys. Rev. E {\bf 55} (1996) 650
[cond-mat/9507080].}
\lref\Clem{J.R.~Clem, Phys. Rev. Lett. {\bf 38} (1977) 1425.  See also E.H.~Brandt,
J. Low Temp. Phys. {\bf 44} (1981) 33; 59.}
\lref\KL{R.D.~Kamien and T.C.~Lubensky, J. Phys. I France {\bf 3} (1993) 2131.}
\lref\TARMEY{V.~G.\ Taratuta and R.~B.\ Meyer, Liquid Crystals {\bf 2},
373 (1987).}
\lref\Stern{D.R.~Nelson and A.~Stern, to appear in {\sl Proceedings of the XIV Sitges
Conference, ``Complex Behavior of Glassy Systems'', June 10-14, 1996}, edited by
M.~Rubi (1996) [cond-mat/9701001].}
\lref\NelSeu{D.R.~Nelson, Phys. Rev. Lett. {\bf60} (1988) 1973;
D.R.~Nelson and H.S. Seung, Phys. Rev. B {\bf 39} (1989) 9153.}
\lref\Leon{L.~Balents and D.R.~Nelson, Phys. Rev. Lett. {\bf 73} (1994)
2618 [cond-mat/9406028]; Phys. Rev. B {\bf 52} (1995)
12951 [cond-mat/9503084].}
\lref\Hatano{N.~Hatano and D.R.~Nelson, Phys. Rev. Lett. {\bf 77} (1996) 570
[cond-mat/9603165].}
\lref\Stern{D.R.~Nelson and A.~Stern, to appear in {\sl Proceedings of the XIV Sitges
Conference, ``Complex Behavior of Glassy Systems'', June 10-14, 1996}, edited by
M.~Rubi (1996) [cond-mat/9701001].}
\lref\MN{M.C.~Marchetti and D.R.~Nelson, Phys. Rev. B {\bf 41} (1990) 1910.}
\lref\Danna{M.V.~Indenbom, C.J.~van der Beek, V.~Berseth, W.~Benoit, G.~D'Anna,
A.~Erb, E.~Walker and R.~Flukiger, Nature {\bf 385} (1997) 702.}
\lref\DRN{D.R.~Nelson, Phys. Rev. Lett. {\bf60}, 1973 (1988);
D.R.~Nelson and H.S. Seung, Phys. Rev. B {\bf 39}, 9153 (1989).}
\lref\LL{L. Landau and E. Lifshitz, {\sl Theory of Elasticity}
3$^{rd}$ ed.  (Pergamon, 1986).}
\Title{}{\vbox{\centerline{Force Free Configurations of Vortices in High
Temperature}\vskip2pt
\centerline{Superconductors near the Melting Transition}}}

\centerline{
Randall D. Kamien\footnote{$^\dagger$}{email: {\tt kamien@dept.physics.upenn.edu}}}
\centerline{\sl Department of Physics and
Astronomy, University of Pennsylvania, Philadelphia,
PA 19104}

\vskip .3in
We study force-free configurations of Abrikosov flux lines in the line-liquid
and line-crystal limit, near the melting transition at $H_{m}$.  We show that the 
condition for 
zero force configurations can be solved by appealing to the structure
of chiral liquid crystalline phases.
\Date{26 April 1998; revised 15 June 1998}

In high temperature superconductors, the extremely large ratio of the London
penetration depth $\lambda$ to the coherence length $\xi$ suggests that
the most important degrees of freedom are Abrikosov vortex excitations.
The configuration of flux lines in applied currents and fields thus
becomes of great interest.  It is therefore useful to construct a theory
of the flux-lines themselves which may be used to study their conformations.

We start by considering the London equation for a superconductor, which
relates the current density $\bf j$ to the magnetic field ${\bf B}$:
\eqn\elondon{\curl {\bf j} = - {c\over 4\pi\lambda^2} {\bf B}}
where $\lambda$ is the London penetration depth.  This is, of course,
supplemented by Maxwell's equation:
\eqn\emaxwell{\curl {\bf B} = {4\pi\over c} {\bf j}.}
These two equations predict much of the phenomenology of superconductors.  
In particular, if ${\bf B}$ is along the $\hat z$-axis, then the London
equation predicts that screening currents will circulate in the $xy$-plane.  
If the magnetic field is confined within flux-tubes parallel to the $\hat z$-axis
then the currents will circulate about these confined regions.  These flux tubes
will form the Abrikosov flux-line lattice.

Under an applied current, flux-lines will adopt a steady
state configuration in which there is no net force.  
In the absence of pinning the Lorentz force per unit length on
the vortices, ${\bf F}/\ell =  {\bf B}\times {\bf j}$, will
be balanced by repulsive vortex--vortex interactions.  
In a ``force-free'' configuration \ref\CE{A.M.~Campbell and
J.E.~Evetts, Advan. Phys. {\bf 21} (1972) 199.}\ it is necessary that 
${\bf j}$ be parallel to ${\bf B}$:   
\eqn\enoforce{\alpha({\bf x}){\bf B}({\bf x}) = {c\over 4\pi}{\bf j}({\bf x}) = 
\curl{\bf B}({\bf x})}
where the last equality follows from Maxwell's equation and $\alpha({\bf x})$
is a spatially-varying scalar.  In a superconductor,
the magnetic field is confined to be near flux tubes and along their
tangents.  If the flux-lines trace out the curves ${\bf R}_i(s)$ where
$s$ is their arclength, then
\eqn\etangent{{\bf m}({\bf x}) = \int ds\sum_i {d{\bf R_i}\over ds}\delta^3
\left({\bf R_i}(s) - {\bf x}\right)}
is the local tangent density of flux-lines.  Since flux-lines
cannot end we have 
\eqn\enoends{\nabla\dot{\bf m}=0.}  
Using standard techniques for treating topological defects \LL\ we have:
\eqn\elon{\left[1-\lambda^2\nabla^2\right]{\bf B}=\Phi_0 {\bf m}}  
where $\Phi_0$ is the flux quantum.
It is useful to decompose ${\bf m}$ as 
the product of a unit vector $\bf\hat n$ and an areal density $\rho$,
${\bf m}= \rho {\bf \hat n}$ \KL .  
If $\alpha$ varies on a lengthscale long compared with the penetration
depth then by applying the operator $\left[1-\lambda^2\nabla^2\right]$ to \enoforce\
we find:
\eqn\ealphaconstant{\nabla\rho\times{\bf\hat n} + \rho\curl{\bf\hat n} \approx 
\alpha\rho{\bf\hat n}.}
If we consider the system near $H_{m}$ where the flux-lines are dense,
we can take $\rho\approx \rho_0$, a constant.  In this case \enoends\ and
\ealphaconstant\ become:
\eqn\eliqcry{\eqalign{\nabla\dot{\bf\hat n}&=0\cr
\curl{\bf\hat n} &=\alpha{\bf\hat n}\cr}}
These two equations are familiar in the field of liquid crystals: the flux-lines
will attempt to adopt a configuration with no splay ($\nabla\dot{\bf\hat n}=0$), 
no bend (${\bf\hat n}\times[\curl{\bf\hat n}]=0$), but with twist
${\bf\hat n}\dot\curl{\bf\hat n}=\alpha$.  
We will pursue this analogy
with liquid crystals.  Of course, the flux-line density does not need to
be uniform.  The liquid crystal analogy will allow us, however, to 
consider a class of paradigmatic vortex configurations which do not
require density variations and are thus of low energy.  

For simplicity,
we consider a superconductor in a magnetic field, applied along 
the $z$-axis.  The Abrikosov flux lattice can be modeled as an
elastic medium \MN :
\eqn\eelas{F_{\rm Lattice} = {1\over 2}\int d^3\!x\,\left\{ c_{11}u_{ii}^2 + 2c_{66} \left[u_{ij}^2-
u_{ii}^2\right] 
+ c_{44}\left(\partial_z
\vec u\right)^2\right\},}
where $\vec u$ is the two-dimensional displacement vector (perpendicular to the
average flux line direction), $u_{ij}$ is the two-dimensional strain tensor
$u_{ij}=(\partial_iu_j+\partial_ju_i)/2$ and we have used
the elastic constants $c_{ij}$ as defined in \ref\BrandtII{E.H.~Brandt and U.~Essman, Phys.
Status Solidi B {\bf 144} (1987) 13. } .   The equilibrium
conformation will minimize
the elastic free energy while maintaining a force-free configuration.

First we consider the case just above $H_{m}$ where
$c_{66}$ vanishes, the flux-liquid \DRN .
In this case, when the flux-lines are aligned by an
external magnetic field, we can directly show that a current parallel to the
field tends to twist the flux-line tangents, as in a cholesteric liquid crystal.
We employ the duality mapping between the superfluid and the superconductor
\ref\Peskin{M.E.~Peskin, Ann. Phys. {\bf 113} (1978) 122; see also M.P.A.~Fisher and
D.-H.~Lee, Phys. Rev. B {\bf 39} (1989) 2756.}\ under an applied local current.
We write ${\bf j}^0 = \rho_ee{\bf v}^0$ where $\bf v$ is the Cooper-pair velocity, 
$\rho_e$ is the
pair density and $e$ is the pair charge. The partition
function for the London
theory in an applied field $\bf H$ is:
\eqn\ai{
Z=\int [d\vb] [d\vA]\,\delta[\nabla\dot\vA]\exp\left\{-\int d^3\!x\,{m\rho_e e^2\over 2}
(\vb-\vb^0-\vA)^2 + \half(\curl\vA)^2
+\vH\dot\curl\vA\right\},}
where $m$ is the mass of the Cooper-pair.
Writing the velocity in Fourier space in terms of longitudinal and transverse components,
\eqn\aii{\vb({\bf k}) = i{\bf k}\phi + {i{\bf k}\!\times\!\vm\over \vert
{\bf k}\vert^2}}
where $\vm$ is the density of flux-vortices pointing in the $\hat\vm$ direction.
Since flux-lines cannot begin or end in the sample $\nabla\dot\vm=0$.
Upon substituting \aii\ into \ai\ and integrating out $\phi$ and $\vA$ 
we have
(to leading order in momentum)
\eqn\avi{Z=\int [d\vm]\delta[\nabla\dot\vm]\exp\left\{-\int d^3\!x\,\left[\eqalign{
&\half \vm^2 - \vm\dot\vH -\vb^0\dot\nabla\times\vH 
+\vb^0\dot\nabla\times\vm    \cr}\right]\right\}}
where we have omitted terms independent of $\vm$ and
$\vb$ .    The first two terms are responsible for
the presence of vortices in the superconductor -- they favor
$\vm=\vH$.  The next term induces the screening current in the Meissner
phase.  The last term is the new interaction which tends to twist the
vortices around the applied current.  If ${\bf H}$ is along $\hat z$, it
is natural to write ${\bf m} \approx \rho{\bf\hat z} + \rho_0\delta\vec n$
where $\delta\vec n$ is the projection of the average tangent onto the
$xy$-plane.  Then $\nabla\dot{\bf m}$ becomes:
\eqn\econservv{\partial_z\delta \rho + \rho_0\grad\dot\delta\vec n =0.}
This constraint can be solved \TARMEY\ by introducing a two-dimensional
vector field $\vec u$ and writing $\delta\rho = -\rho_0\grad\dot\vec u$ and
$\delta\vec n =\partial_z\vec u$.  
In terms of this field $\vec u$
\eqn\etotale{F_{\rm total} =  \int d^3\!x\,\left\{
{c_{11}\over 2}u_{ii}^2 + {c_{44}\over 2}\left(
\partial_z\vec u\right)^2-
{m\rho_0\over 2e^2\rho_c}j
\dot{\grad\times\partial_z\vec u}\right\},}
where we have allowed for anisotropic elastic constants.     
This theory is simply the theory of polymer cholesterics, in the limit
of small pitch \KL .  For large deviations one might expect that the
flux-lines will rotate in a plane perpendicular to a pitch direction.  This
configuration was, in fact, proposed in the seminal work of Campbell and Evetts \CE .
Moreover, an additional conformation is possible in a finite radius, 
cylindrical sample,
namely a double-twist configuration as in the blue-phase of chiral liquid crystals.
This possible double-twist configuration is shown in figure 1.  In this double-twist
conformation the $\bf B$-field and $\bf j$ wrap around each other, simultaneously
satisfying the Maxwell and London equations \emaxwell\ and \elondon .  It
was correctly noted in \CE\ that this configuration would be energetically
unacceptable as the radius of the cylinder grew -- the flux lines on the boundary
of the sample would grow unacceptably long.  However, as we shall see in
the following,
a defect-riddled state can allow local configurations similar to those
shown in figure 1, with finite displacements of the flux-lines.  The
handedness of the rotation of the flux-lines is determined by the right-hand-rule
and the direction of the current.  Note, however, that an equally acceptable
force-free conformation would be {\sl absolutely} straight flux-lines parallel
to the applied current.  The difficulty with this is that thermal fluctuations
will destabilize this state and lead to a helical instability of flux-lines,
as predicted by Clem \Clem\ in 1977.  As a consistency check, we note that 
if we were to consider the effect of the Lorentz forces acting 
on the individual flux-lines that there is an instability also at any 
finite current \ref\leo{We thank L.~Radzihovsky for discussions on this point.}\
towards helical flux-line trajectories.  

We note that there is a certain duality between the current and the magnetic field
in the London-Maxwell equations.  In particular, the equations are
invariant under
\eqn\edualone{\eqalign{{\bf j} &\rightarrow {c\over 4\pi\lambda}{\bf B}\cr
{\bf B} &\rightarrow -{4\pi\lambda\over c} {\bf j}.\cr}}
It would thus be natural to consider the dual physical situation to the 
Abrikosov flux-lattice.  In this case, the current would flow along the $\hat z$-axis
leading to a screening magnetic circulation in the $xy$-plane.  If the
current were confined into regions, so would be the circulating magnetic field.  
Physically, this is accomplished via flux lines tracing out helical
trajectories: the $xy$-components of the flux-line tangents circulate
in that plane, dragging the magnetic field with them providing the
necessary magnetic field.

There is, however, an essential difference between the Abrikosov solution
and its dual: in the original problem, quantum mechanics imposes a constraint
on the amount of magnetic flux that could be confined in a a flux-tube -- single-
valuedness of the wavefunction implies that the flux must be 
an integer multiple of the flux quantum $\Phi_0 = {2\pi\hbar c\over 2e}$.  
This constraint is responsible for the presence of a second-order transition
between the Meissner state and the Abrikosov state.  In the dual case, there
is no equivalent quantization of current flux.  It is easy to understand why in
the dual language: a helical flux-line can execute an arbitrarily long-pitched
wobble which allows $[\curl{\bf B}]_\perpp$ to be arbitrarily small.   This is what
allows
Clem's helical instability.  

When $c_{66}\ne 0$ we are forced to consider a crystalline structure with 
a force-free conformation of the flux-lines.   This problem has been
considered in the context of liquid crystals -- namely, how a chiral
line-crystal minimizes its free-energy in the presence of the two-competing
tendencies to twist and to have periodic order.  These two tendencies
frustrate each other and thus, as in the Renn-Lubensky twist-grain-boundary (TGB)
phase of smectic liquid crystals the frustration will be resolved via
the introduction of topological defects -- screw dislocations.  It is amusing that the TGB phase
is the analog of the Abrikosov phase of the superconductor and so in the problem
at hand we are minimizing the stresses of the real Abrikosov lattice with a
``dual'' Abrikosov lattice. 
  
In \KNMoire\ two types of crystal defect arrays were considered.  One array consisted
of a periodic arrangement of tilt-grain-boundaries (TGB) which would change
the local flux-line direction.  The other array was made of 
helicoidal-grain-boundaries (HCB),
each of which is a honeycomb lattice of screw-dislocations lying in the $xy$-plane.  
A single isolated HCB leads to a twisting of the crystalline order along the
flux-line direction.     
If we were to consider stacking many HCB's together with some spacing $d'$, this
twisted moir\'e state would have both twisting of $\delta\vec n$ as well as twisting
of the crystal directions.  This is similar to the physics of blue phases in
chiral liquid crystals \ref\MW{S.~Meiboom, J.P.~Sethna, P.W.~Anderson and 
W.F.~Brinkman, Phys. Rev. Lett. {\bf 46},
1216 (1981); 
D.C.~Wright and N.D.~Mermin, Rev. Mod. Phys. {\bf 61}, 385 (1989).}.  
In chiral liquid crystals, there is
a tendency for the local director
$\bf n$ to twist.  However, in blue phases this twist manifests itself
in double-twist cylinders.  Taking the nematic director field as a local
tangent vector density for lines, these double-twist cylinders become
rope-like bundles of twisted lines.  Analogously, a twisted bundle of 
flux-lines will allow the magnetic field to circulate while keeping the
flux-lines, on average along a single direction.  
While in the softer liquid crystal theory the 
elastic energy cost of this deformation is proportional to the angle
of rotation \KNMoire , in the flux-line system
interactions between the screw dislocations of the
vortex lattice will lead 
to logarithmic corrections to this energy.   In any event, the energy
of a grain boundary per unit area will be finite.
We propose these defected states as paradigms for 
a flux-line lattice under an applied, parallel current.  

Notice that we can have no twist if $\alpha=0$.  However, ${\bf j}=\alpha {\bf B}$.
If current flows through the superconductor then it must flow on the oft-neglected
boundaries of the sample.  Thus,
to study the energetics of this state, we 
must include the usual London energy for the supercurrent.
We consider a current along the magnetic field direction $\hat z$ and
assume that there is a flux-line lattice.
If $v$ is the Cooper-pair velocity and $\rho_c$ is the density
of Cooper-pairs, then the total current is $I_z=je\rho_cAv$ where 
$A$ is the cross sectional area of the region in which
current flows.  If there are no defects in the
flux lattice the current must flow within a penetration depth $\lambda$
of the sample boundary and $A\approx 2\pi R\lambda$.  
The London energy for this current configuration is, per unit length along 
$\hat z$,
\eqn\elondon{F/L = 2\pi\int_{R-\lambda}^R rdr m_e\rho_c v^2
= {m_e\over 4\pi R\lambda\rho_c e^2}I_z^2,}
where $m_e$ is the mass of the electron.

If we allow the Abrikosov lattice to have defects then the current can
flow through more of the cross section thus lowering the London energy.  Of
course,  the energy decrease will be offset by the energy of the screw dislocations
in the flux-line lattice.
If we consider a moir\'e configuration which is reasonably dense then
as a rough approximation we take
$\curl{\delta\vec n}$ to be uniform along $\hat z$ .  
This implies that the
current runs uniformly 
through the entire cross-section of the sample.  The decreased London
energy is:
\eqn\elondonn{F_{\rm defects}/L = {m\over  \pi R^2\rho_c e^2}I_z^2.}
In the moir\'e state we must add the energy of the dislocation lattice that
produces the twisted configuration.  In a crystalline lattice the
energy in the strain field due to a
single dislocation diverges logarithmically with system
size.  If we have a network of dislocations as shown in figure 2, 
however, the strain energy of the
lattice is finite.  In this case the largest contribution to the
energy of a grain-boundary is due to the logarithmic interactions of the
screw-dislocations.  
We take the energy per unit
length of a screw-dislocation to 
be $\epsilon_0\ln (d/\zeta)$ where
$d$ is the average defect spacing and $\zeta$ is the defect core size \ref\rfoot{
We reserve $\xi$ for the coherence length, the size of the Abrikosov vortices.}.
The energy cost per unit length along $\hat z$ of the
sample is therefore:
\eqn\eed{E_{\rm defects}/L = \epsilon_0 \ln(d/\zeta) {\pi R^2\over d^2}.}  
With a uniform current density 
Maxwell's equation gives $\nabla\!\times\!{\bf B} = 4I_z/(c R^2)$.
In turn,
the defect density is determined by the amount of $\curl\delta\vec n$ required
to produce the uniform current.  For a defect spacing $d$ we estimate \ref\foottwo{
There is a technical issue at this point which we are simplifying.  A sum rule
\KNMoire\ relates the curl of $\delta\vec n$ to the rate at which the bond-angle
$\theta_6$ rotates, $\partial_z\grad\cross\vec u - \grad\cross
\partial_z\vec u= \alpha$ where $\alpha$ is the screw-dislocation density.  
Any particular dislocation complexion will lead to both twist and bond-angle
rotation -- the details of the geometry will determine the ratio of
the two effects. In \elind\
we have assumed that the defect strain is shared equally
between the two deformations.  A more careful 
calculation should only change our estimate by a factor of ${\cal O}(1)$
in any non-pathological geometry.}
\eqn\elind{\grad\times{\delta\vec n} \approx {a_0\over 2d^2}.}
Putting
this together with Maxwell's equation, we get
\eqn\esoll{{\pi R^2\over d^2} = {8\pi I_z\over c}{1\over\sqrt{\Phi_0H_z}}}  
where we have used $\rho_0=H_z/\Phi_0$ and $a_0=1/\sqrt{\rho_0}$.
Thus the energy cost from the defects is:
\eqn\eedi{E_{\rm defects}/L = \ln\left[{c\sqrt{\Phi_0H_z}\pi R^2\over
8\pi I_z\zeta^2}\right] {4\pi\epsilon_0I_z\over c\sqrt{\Phi_0H_z}}.}

Putting together all the energies we may compare the energy of the
moir\'e structure with that of the untwisted Abrikosov structure.  There
will be an instability towards a twisted state when
\eqn\eequate{{m\over 2\pi R\lambda\rho_ce^2}I_z^2\ge
{m\over \pi R^2\rho_ce^2}I_z^2 + E_{\rm defects}/L.} 
If $\lambda\ll R$ we may neglect the first term on the right hand side
of \eequate\ and find that the moir\'e state is favored when 
\eqn\izz{I_z\ge \ln\left[{c\sqrt{\Phi_0H_z}\pi R^2\over
8\pi I_z\zeta^2}\right]{8\pi^2 R\lambda\rho_ce^2\epsilon_0\over m_ec\sqrt{\Phi_0
H_z}},}
or when the applied current density, $j_z=I_z/(\pi R^2)$ is
\eqn\ejz{j_z\ge {\lambda\over R}\ln\left[{c\sqrt{\Phi_0H_z}\over
8\pi j_z\zeta^2}\right]{8\pi\rho_ce^2\epsilon_0\over m_ec\sqrt{\Phi_0
H_z}}.}
Thus, as the system size increases the current density necessary
to go to the moir\'e state goes to zero.  Thus the instability of 
Clem may be stabilized through the lattice structure and its screw
dislocations.
The moir\'e state that proposed here is not exactly force-free. In each
cell of the honeycomb, the flux-line displacement is a $z$-dependent rotation:
$u_i = Cz\epsilon_{ij}x_j$.  This configuration has $\curl{\bf B}$ parallel
to $\bf\hat z$, {\sl not} the local $\bf B$.  However, when averaged
over one cell, $\bf B$ is parallel to $\bf\hat z$ and so there is no
net force on each bundle of vortices.  It is clear, however, that small adjustments
to the flux-line locations that do not change their topology 
can yield an entirely force-free configuration.  

We end by commenting on recent experiments \Danna\ performed on 
thin-film,\break YBa$_2$Cu$_3$O$_{7-\delta}$, 
$ab$-plane superconductors which have seen what are called ``vortex''-twisters, 
composed of thousands of flux-lines.  These twisters are formed by first
applying a magnetic field
along the long-direction 
of the sample creating an Abrikosov flux-line-lattice.
An additional field $H_\perp$ is applied perpendicular to the plane of the
flux-lines.  As usual, the superconductor responds with a screening current which  
generates Lorentz forces on the vortices.  Because of the anisotropy
of the superconductor the flux-lines will remain, for the most part, in the $ab$-plane
and will not realign with the net magnetic field direction if $H_\perp$ is small
enough \Leon .  On the short side of the sample these Lorentz forces push the
flux lines along the direction either parallel or anti-parallel to the $c$-axis depending
on the sign of the force.  This bending will only occur near the sample edge
as there is a competition for each flux-line 
between the in-plane pinning and the tendency to align
with the net field \Hatano .  The forces build up and the Bean critical state \ref\Bean{
See, for instance, P.-G.~de~Gennes, {\sl Superconductivity of Metals and Alloys}, 
(Benjamin, New York, 1966).}\ is formed.  
For current to flow along the sides of the superconductor, the flux-lines must
twist about, giving a non-zero $\curl{\bf B}$.
Moreover, 
these vortex
twisters can be made ever more stable and compact through ``work-hardening'' via
an AC component of $H_\perp$.  When the AC field is 
removed, the work-hardened bundles are stable for hours, while the
unhardened bundles are not stable at all.
Since the moir\'e state is a highly-ordered structure, it is unlikely that these
vortex-twisters form a complete lattice of honeycomb-boundaries.  However,
it is possible that the local structure of these vortices resembles the
highly entangled moir\'e state.  This could explain the work-hardening:  when
there are many screw dislocations it is difficult for the flux-line lattice
to relax, as the defects cannot cross the flux-lines without cutting them.
A flux bundle twisted as in figure 1 could relax easily as there are no
topological impedements.

Along these lines, it would be interesting to study (experimentally
and theoretically) the dynamics of the flux-line lattice under
an applied current.  In particular, just above $H_{m}$, the flux-line
picture would suggest that there should be a striking frequency dependence
of the $I-V$ curve for currents along the field axis.  If the frequency
is smaller than a typical vortex diffusion time then the 
flux-lines can adjust their configurations to allow for
current flow.  However, as the frequency is increased the flux-lines
will not be able to adjust for the current.  This is reminiscent
of the visco-elastic behavior of polymer melts.  Moreover, if work-hardening
of the Abrikosov lattice is possible, one might be able to increase the
critical current in the sample \ref\drnn{D.R.~Nelson, Nature {\bf 385} (1997) 675.}.
These possibilities would verify the dual picture of high-$T_C$ superconductors
which focuses on the Abrikosov flux-lines.

It is a pleasure to acknowledge stimulating discussions with J.R.~Clem,
T.C.~Lubensky, M.C. Marchetti, L.~Radzihovsky and especially D.R.~Nelson.  
We also thank the Aspen Center for Theoretical Physics, where some
of this work was done.
This work was supported, in part, by an award from Research Corporation and an
NSF Career Award through Grant Number DMR97-32963.
\nfig\fone{Configuration of flux-lines and current in a wire (both follow
the heavy lines).  The
flux-lines are parallel to the current everywhere and both wrap around
the center of the wire.  Note that there is a nonvanishing $\curl{\bf B}$ and
$\curl{\bf j}$.  The applied field $\bf H$ and the average current density 
$\overline
{\bf j}$ are parallel.}
\nfig\ftwo{Proposed braided (moir\'e) state of flux-lines.  The dark lines
make up the honeycomb network of screw dislocations.  This texture has an
approximately uniform $\curl{\bf B}$ and so the current will flow
uniformly through the cross-section of the sample.}
\listrefs
\listfigs

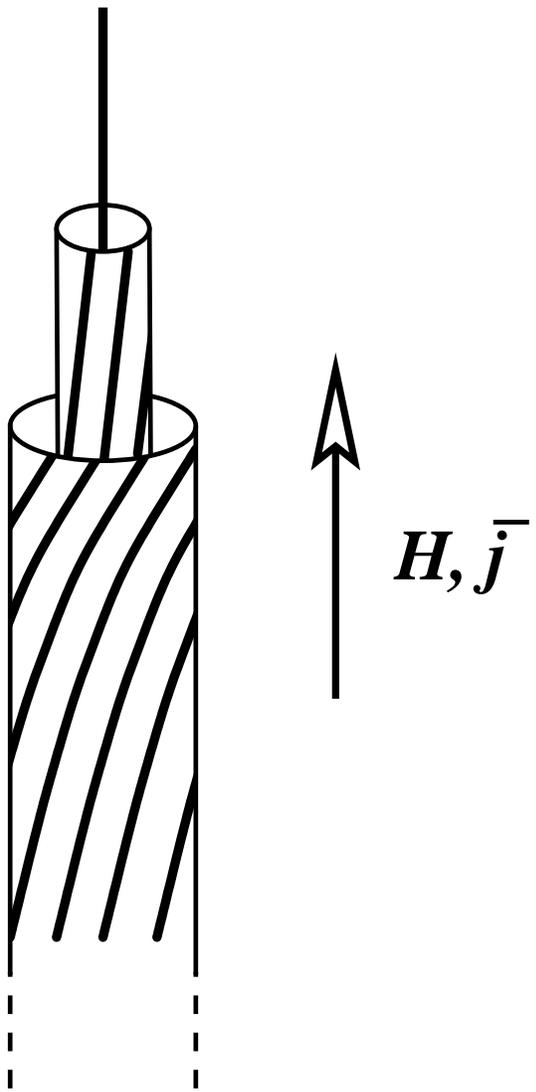
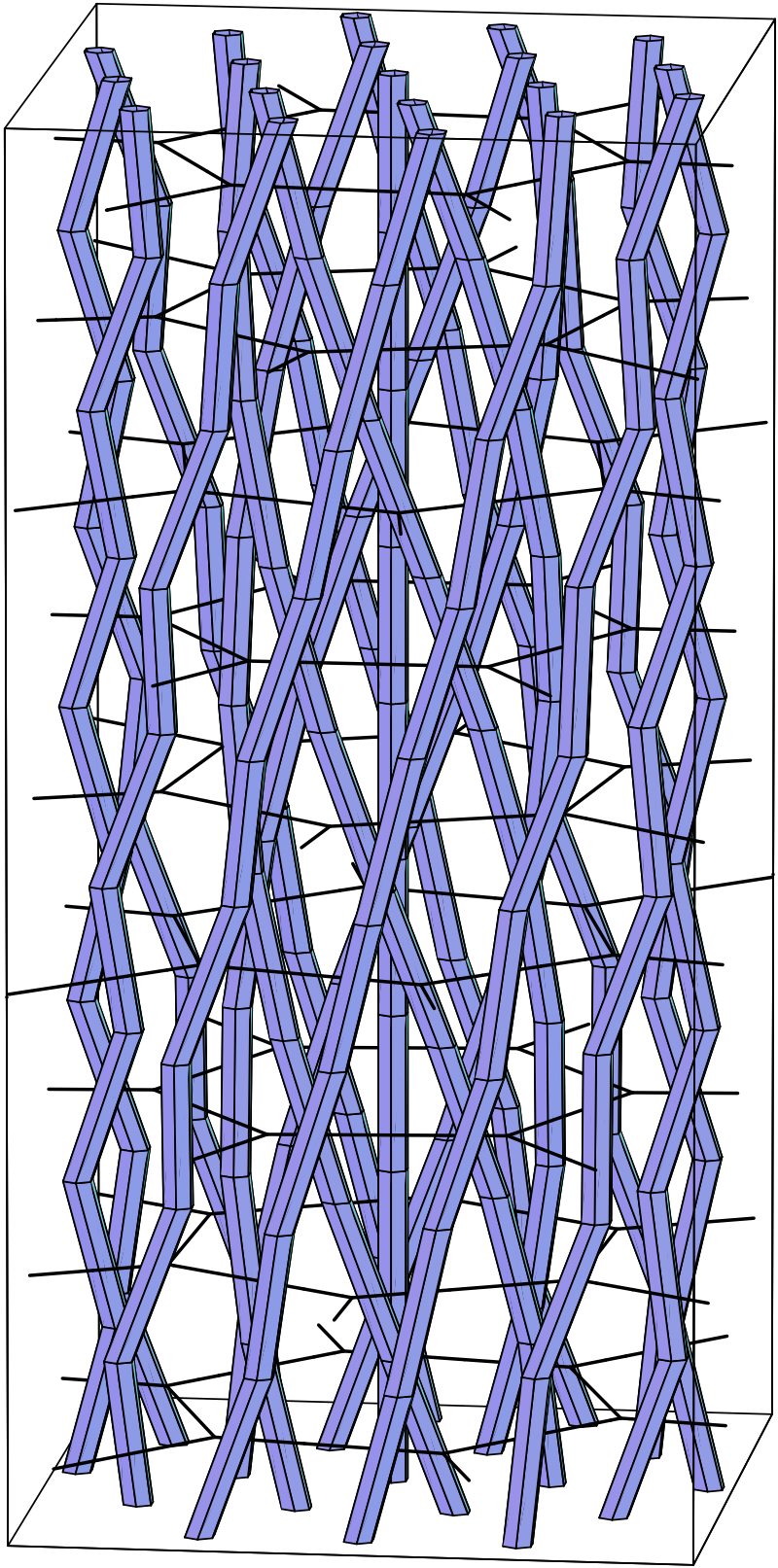
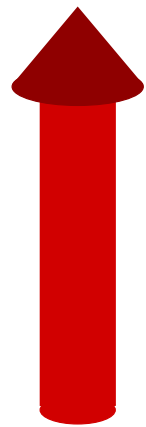
\bye